\documentclass[aps,prd,onecolumn,notitlepage,preprint,showpacs, showkeys,superscriptaddress]{revtex4-2}

\usepackage{graphicx}
\usepackage{epstopdf}
\usepackage{amsmath}
\usepackage[section]{placeins}
\usepackage{slashed}
\usepackage{color}
\usepackage{soul}
\usepackage{appendix}
\usepackage{placeins}
\usepackage[utf8]{inputenc}
\usepackage{pstricks}
\usepackage{multirow}
\usepackage{hyperref}
\usepackage{subfigure}
\usepackage{epsfig}
\usepackage{cancel}
\usepackage{braket}
\usepackage[normalem]{ulem} 
\usepackage{longtable,booktabs}
\newcommand{\be}{\begin{equation}}
\newcommand{\ee}{\end{equation}}
\newcommand{\ben}{\begin{eqnarray}}
\newcommand{\een}{\end{eqnarray}}

\begin{document}

\title{ $D^*\pi$ interaction from the lineshape of $D_1(2420)$ in $B$-decays}

\author{Pedro Brandão}
\email[]{pedro.brandao@ufba.br}
\affiliation{Instituto de Física, Universidade Federal da Bahia, Campus Ondina, Salvador, Bahia 40170-115, Brazil}

\author{Breno Agatão}
\email[]{bgarcia@if.usp.br}
\affiliation{Instituto de F\'{\i}sica, Universidade de S\~{a}o Paulo, Rua do Mat\~{a}o, São Paulo SP, 05508-090, Brazil}   

\author{Luciano M. Abreu}
\email[]{luciano.abreu@ufba.br}
\affiliation{Instituto de Física, Universidade Federal da Bahia, Campus Ondina, Salvador, Bahia 40170-115, Brazil}

\author{K.~P.~Khemchandani}
\email[]{kanchan.khemchandani@unifesp.br}
\affiliation{Universidade Federal de S\~ao Paulo, C.P. 01302-907, S\~ao Paulo, Brazil}

\author{A.~Mart\'inez~Torres}
\email[]{amartine@if.usp.br}
\affiliation{Instituto de F\'{\i}sica, Universidade de S\~{a}o Paulo, Rua do Mat\~{a}o, São Paulo SP, 05508-090, Brazil}

\vspace{1cm}

\begin{abstract}
We present a model calculation to reproduce the differential mass distribution for the $D^*\pi$ system in the reactions $B^- \to D^{*+} \pi^- \pi^-$ and $B^{+}\to D_s^+D^{*-}\pi^{+}$ analyzed by the LHCb Collaboration, which shows a dominant signal for $D_1(2420)$. 

We consider a model based on coupled channel meson-meson interactions that can describe the properties of $D_1(2420)$ in terms of the underlying dynamics, use it to determine the invariant mass distribution of the $D^*\pi$ system, and compare the results with the experimental data. We also determine the $D^*\pi$ scattering length, for which different values are available from different sources, leading to a controversy. To our knowledge, this is the first attempt to reproduce the mentioned data using model calculations. Our formalism relies on the hadronization of different mesons through a weak decay, allowing for the final-state (strong) interactions among the relevant constituents. We benefit from our previous work when obtaining the amplitudes corresponding to the strong interactions. We hope that our findings can be useful in further investigations of the properties of $D_1(2420)$.

\end{abstract}

\maketitle
\newpage
\section{\label{sec:level1} Introduction}
It is well-known that Goldstone bosons are related to the spontaneous breaking of an underlying continuous symmetry. For instance, in hadron physics, the pion is related to the spontaneous breaking of the chiral symmetry. Hence, understanding its interaction with hadrons can lead to gaining important insights into the non-perturbative regime of the theory of strong interactions. One such system, which has received increasing interest in current times, is that of $D^*\pi$. For example, data on the momentum correlation function between $D^*$ and $\pi$ was recently published by the Alice Collaboration~\cite{ALICE:2024bhk}. The same work determined the value of the $D^*\pi$ scattering length, which was found to have an absolute value smaller than 0.1 fm. This value is at odds with those determined by models based on unitarized chiral perturbation theory~\cite{Guo:2018tjx,Guo:2009ct,Abreu:2011ic,Geng:2010vw}, which include studies that fit the model parameters using data from lattice QCD~\cite{Guo:2018tjx,Guo:2009ct}. 
In such a situation, recent publications on experimental data for $B$ decaying to final states containing $D^*$ and $\pi$ may be considered as an excellent alternative source of information on the system. Specifically, LHCb reported data on $B^{+}\to D_s^+D^{*-}\pi^{+}$~\cite{LHCb:2024vhs}, where the most important strong interactions are expected to come from the $D^{*-}\pi^+$ system, since only these mesons have light-flavor quarks in common.
It is useful to recall that data from the same collaboration are also available on the process $B^{-}\to D^{*+}\pi^{-}\pi^-$~\cite{LHCb:2019juy}. The purpose of this paper is to build a model to study these weak decay processes, including strong final-state interactions among different mesons whenever relevant.

It is important to mention that the present work is also motivated by the discussions initiated in our previous work~\cite{Khemchandani:2023xup}, where we investigated the relation between light axial mesons with charm, which are $D_1(2420)$ and $D_1(2430)$, and two-meson coupled channel interactions. We discussed that even though the existence of two axial mesons with charm, having similar masses and different widths, could be expected from the arguments of heavy-quark symmetry, their properties cannot easily be described within models based on quark/hadron dynamics. For example, attempts made by considering different mixing angles between the two states show that it is difficult to obtain a good description of the properties of these mesons~\cite{Ferretti:2015rsa,Cahn:2003cw,Zhou:2011sp,Ni:2021pce}. Also, the works in Refs.~\cite{Zhou:2011sp,Khemchandani:2023xup,Malabarba:2022pdo,Gamermann:2007fi} suggest that hadron loops or two-meson coupled-channel interactions play an important role in describing the properties of the light axial mesons.

The formalism used here is based on earlier studies~\cite{Malabarba:2022pdo,Gamermann:2007fi}, in which coupled channel $S$-wave interactions among the following systems are considered: $D^*\pi$, $D\rho$, $\bar K D_s^*$, $D_s\bar K^*$, $\eta D^*$, $D\omega$, $\eta_c D^*$, $D J/\Psi$, $D\phi$, and $\eta^\prime D^*$. The interaction kernels in these previous works were obtained from a Lagrangian based on a SU(4) symmetry, which is broken. Such amplitudes are also in agreement with a formalism based on the hidden local symmetry, which considers vector mesons as the gauge bosons~\cite{Bando:1987br}. Here, it is important to mention that the equivalence between the two approaches can be obtained by means of the exchange of light vector mesons in the $t$-channel, which makes use only of the SU(3) part of the SU(4) symmetry considered in Refs.~\cite{Malabarba:2022pdo,Gamermann:2007fi}. Thus, heavy quarks in such interactions are treated as spectators as instructed by the heavy quark symmetry.  As discussed in Ref.~\cite{Khemchandani:2023xup}, two poles arise in such a formalism, the properties of one of which agree well with those of the $D_1(2420)$ known from the experimental data. However, both  mass and width associated with the second pole   are not in good agreement with the properties of the $D_1(2430)$  cataloged by the particle data group~\cite{ParticleDataGroup:2024cfk}. To improve the agreement between the mass and width of the pole found in Ref.~\cite{Malabarba:2022pdo}, a bare quark model pole was added in Ref.~\cite{Khemchandani:2023xup}, which required including a few free parameters in the model. In the present work, though, we keep the model simple, without adding the quark model pole, for a twofold reason. Firstly, our aim here is to reproduce the experimental data for two $B$ decay processes available in Refs.~\cite{LHCb:2024vhs,LHCb:2019juy}, and to extract information on the $D^*\pi$ scattering length. Further, the main feature of the data on the invariant mass distribution of $D^*\pi$ is a clear signal of $D_1(2420)$, while a wide state, such as $D_1(2430)$, is washed out by the background. In such a situation, our main focus is to be able to reproduce the lineshape of $D_1(2420)$ in the $D^*\pi$ invariant mass distribution, which is well described in terms of two-meson dynamics. 
Secondly, the model involves the hadronization of several allowed three-meson systems, two of which interact in the final state to produce a $D^*$ and a pion. As shall be discussed in the subsequent section, the whole formalism is quite cumbersome, requiring regularization of divergent loop integrals, and a simplified final-state interaction minimizes the number of free parameters of the model, especially since the focus is on $D_1(2420)$.

In summary, we construct a model for two processes: $B^- \to D^{*+} \pi^- \pi^-$, and $B^{-}\to D_s^-D^{*+}\pi^{-}$, considering the dominant weak-mechanism to produce such final states. In particular, the quarks produced in the weak process are allowed to hadronize as eight different intermediate states. The mesons produced then interact to produce the desired final state in each process. Some of the main findings of our study are: (1) The final-state interaction among the mesons, considered in the $S$-wave, is found to contribute to the $D^*\pi$ invariant mass more than that expected from the analysis carried out in Refs.~\cite{LHCb:2024vhs,LHCb:2019juy}. The lineshape of the $D^*\pi$ mass distribution is reasonably reproduced without including $D$-wave interactions and other $D$-mesons with mass around 2.4~GeV. Our work does not imply that such contributions may not be necessary to describe the data. We just find that the $S$-wave contribution is larger than the estimate given in Refs.~\cite{LHCb:2024vhs,LHCb:2019juy}.  (2) The scattering length for the $D^*\pi$ system, in isospin 1/2, is found to be more in agreement with the findings of Refs.~\cite{Liu:2012zya}.  The details of the formalism are given in the next section, and in the subsequent section, we present the results of our findings and discussions on them.  


\section{Formalism}\label{sec:level2}

\subsection{The weak vertex and hadronization in $B^{-}\to D_s^-D^{*+}\pi^{-}$ and $ B^-  \to D^{*+}\pi^{-}\pi^{-}$ }\label{subweak}

\begin{figure}[ht!]
    \centering
    \subfigure[]{
    \includegraphics[width=0.48\textwidth]{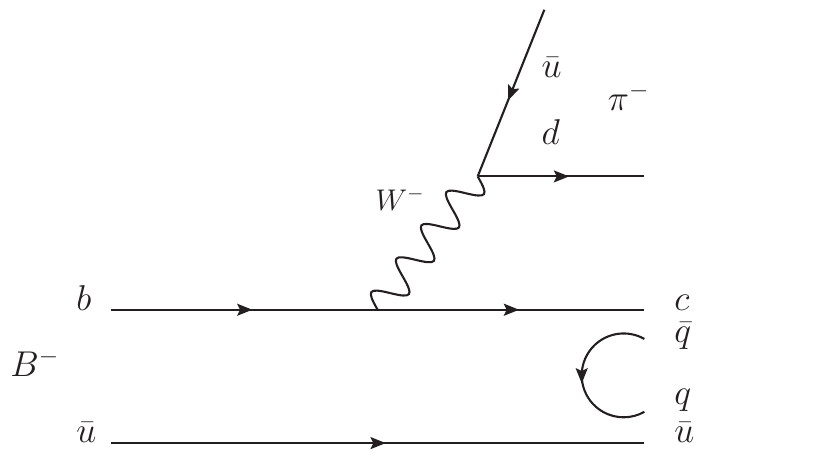}
    \label{quarklv_a}
    }
    \subfigure[]{
    \includegraphics[width=0.48\textwidth]{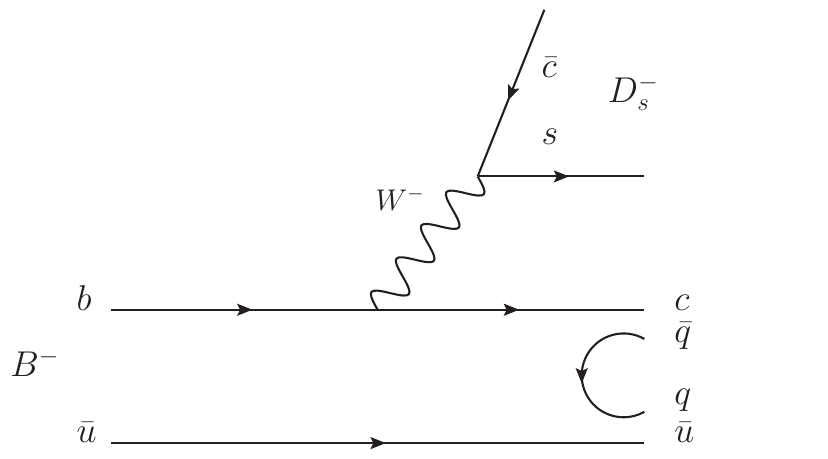}
    \label{quarklv_b}
    }
\caption{Quark-level diagrams with an external emission of a $W^{-}$ boson which contribute to the processes: (a) $B^- \to D^{*+}\pi^{-}\pi^{-}$ and  (b) $B^{-}\to D_s^-D^{*+}\pi^{-}$. Here $q$ denotes a light quark, $q=\{u,d,s\}$. The quark pair generated from the vacuum, together with $c$ and $\bar u$, hadronize as a vector and a pseudoscalar meson which can interact to produce $D^{*+}\pi^-$ eventually. }
\label{quarklv}
\end{figure}

We start by presenting, in Fig.~\ref{quarklv}, the dominant weak transitions for the $B^-\to D^{*+}\pi^{-}\pi^{-}$ and $B^{-}\to D_s^-D^{*+}\pi^{-}$ reactions at the tree level, which proceed through external emission of a $W^-$ boson. 
Here, we should first mention that the latter decay is the conjugate of the process on which the experimental data are available from Ref.~\cite{LHCb:2024vhs}. The choice of treating the conjugate process allows us to present the formalism and the results for the two processes in a uniform way.  This approach is justified because independent fits of the baseline model to the $B^+$ and $B^-$ samples, reported in Ref.~\cite{LHCb:2024vhs}, show statistical agreement and are thus consistent with minimal CP-violating effects.

In the case of $B^{-}\to D^{*+}\pi^{-}\pi^{-}$ [see Fig.~\ref{quarklv_a}], the $b$ quark decays into a $c$ quark by an external emission of a $W^{-}$ boson, which subsequently decays into a $d\bar{u}$ pair to form a $\pi^{-}$ meson. The spectator $\bar{u}$ quark of the reaction, together with a $d\bar{d}$ pair generated with the quantum numbers of the vacuum,  produces the $D^{*+}$ and $\pi^{-}$ mesons in the final state.

The hadronization mechanism producing the mentioned $D^*$ and $\pi$, or other possible mesons, is formally implemented in the formalism as follows: a $\bar{q} q$ pair with the quantum numbers of the vacuum is introduced, where $ \bar{q} q \equiv \sum\limits_{i=\{u,d,s,c\}} \bar{q}_i q_i $ with the $c$ and $\bar u$ available after the weak decay can form directly the final state or an intermediate state consisting of a vector and a pseudoscalar mesons. Mathematically, we have
\begin{equation}
    c\Bar{u}\to\sum_{i=\left\{u,d,s,c\right\}}  c\Bar{q}_{i} q_{i}\Bar{u},
\label{had1}   
\end{equation}
which provides the quark content of all possible intermediate states. 
Further, taking into account that the $(q\bar{q})$-matrix can be written as:
\begin{align}\label{Qmatrice1}
(q\bar{q}) =
\left(
  \begin{array}{cccc}
    u\Bar{u} & u\Bar{d} & u\Bar{s} & u\Bar{c}\\
    d\Bar{u} & d\Bar{d} & d\Bar{s} & d\Bar{c}\\
    s\Bar{u} & s\Bar{d} & s\Bar{s} & s\Bar{c} \\
    c\Bar{u} & c\Bar{d} & c\Bar{s} & c\Bar{c} \\
  \end{array}
\right),
\end{align}
it is easy to see that the structure on the right-hand side of Eq.~($\ref{had1}$) can be obtained by taking the $[(q\bar{q})(q\bar{q})]_{41}$ element. Next, replacing the $(q\bar{q})$ matrix elements in terms of mesons, we get the following representations of the matrices in the $ \text{SU}(4) $ flavor space for the pseudoscalar ($P$) and vector ($V$) mesons,
\begin{align}\label{meson_P}
\centering
q\bar q\Rightarrow P=
\left(
  \begin{array}{cccc}
    \frac{\pi^0}{\sqrt{2}}+\frac{\eta}{\sqrt{3}}+\frac{\eta'}{\sqrt{6}} & \pi^{+} & K^{+} & \bar{D}^{0}\\
    \pi^{-} & -\frac{\pi^0}{\sqrt{2}}+\frac{\eta}{\sqrt{3}}+\frac{\eta'}{\sqrt{6}} & K^0 & D^{-}\\
    K^{-} & \bar{K}^0 & -\frac{\eta}{\sqrt{3}}+\sqrt{\frac{2}{3}}\eta' & D_{s}^{-} \\
    D^{0} & D^{+} & D_{s}^{+} & \eta_{c} \\
  \end{array}
\right),~~~~
\end{align}
\begin{align}\label{meson_V}
\centering
q\bar q \Rightarrow V_{\mu}=
\left(
  \begin{array}{cccc}
    \frac{\rho^0}{\sqrt{2}}+\frac{\omega}{\sqrt{2}} & \rho^{+} & K^{*+} & \Bar{D}^{*0} \\
    \rho^{-} & \frac{-\rho^0}{\sqrt{2}}+\frac{\omega}{\sqrt{2}} & K^{*0} & D^{*-} \\
    K^{*-} & \bar{K}^{*0} & \phi & D_{s}^{*-}\\
    {D}^{*0} & D^{*+} & D_{s}^{*+} & J/\psi \\
  \end{array}
\right)_{\mu}.
\end{align}
In this way, the $[(q\bar{q})(q\bar{q})]_{41}$ element can be written in terms of mesons by considering the matrix elements $(VP)_{41}$ and $(PV)_{41}$. The element $(VP)_{41}$ yields
\begin{equation}
    c\Bar{u}\to \frac{D^{*0}\pi^{0}}{\sqrt{2}} + \frac{D^{*0}\eta}{\sqrt{3}}+ D^{*+}\pi^{-}+D^{*+}_{s}K^{-}  + \cdots, 
\label{had2}
\end{equation}
whereas $(PV)_{41}$ produces
\begin{equation}
    c\Bar{u}\to \frac{D^{0}\rho^{0}}{\sqrt{2}} +\frac{D^0\omega}{\sqrt{2}}+ D^{+}\rho^{-}+D^{+}_{s}K^{*-} + \cdots,
\label{had3}
\end{equation}
where the dots stand for other systems consisting of two heavy mesons, like $\eta_c D^*$, $D J/\Psi$, which would give a negligible contribution, and which we do not consider. 
The combinations obtained in Eqs.~(\ref{had2}) and (\ref{had3}) allow us not only to determine a tree-level background for the decay $B^-\to D^{*+}\pi^{-}\pi^{-} $, but also to calculate the resonant contributions associated with the charmed axial states coming from the rescattering processes (see the discussion in subsequent sections). 

A similar analysis can be applied to the $B^-\to D_s^- D^{*+}\pi^{-}$ decay [see Fig.~\ref{quarklv_b}]. The key difference from the process depicted in Fig.~\ref{quarklv_a} is that the $b$ quark decays into a $ c $ quark via the external emission of a $ W^{-} $ boson, which subsequently produces an $s\bar{c}$ pair to form the $D_s^{-}$ meson. The hadronization of the remaining quarks, again modeled by the production of a virtual $q\bar{q}$ pair, leads to the final state mesons $D^{*+} $ and $ \pi^{-}$, as well as other meson states listed in Eqs.~(\ref{had2}) and~(\ref{had3}), which then rescatter to produce the required final state.

\begin{figure}
\centering
\includegraphics[width=0.5\textwidth]{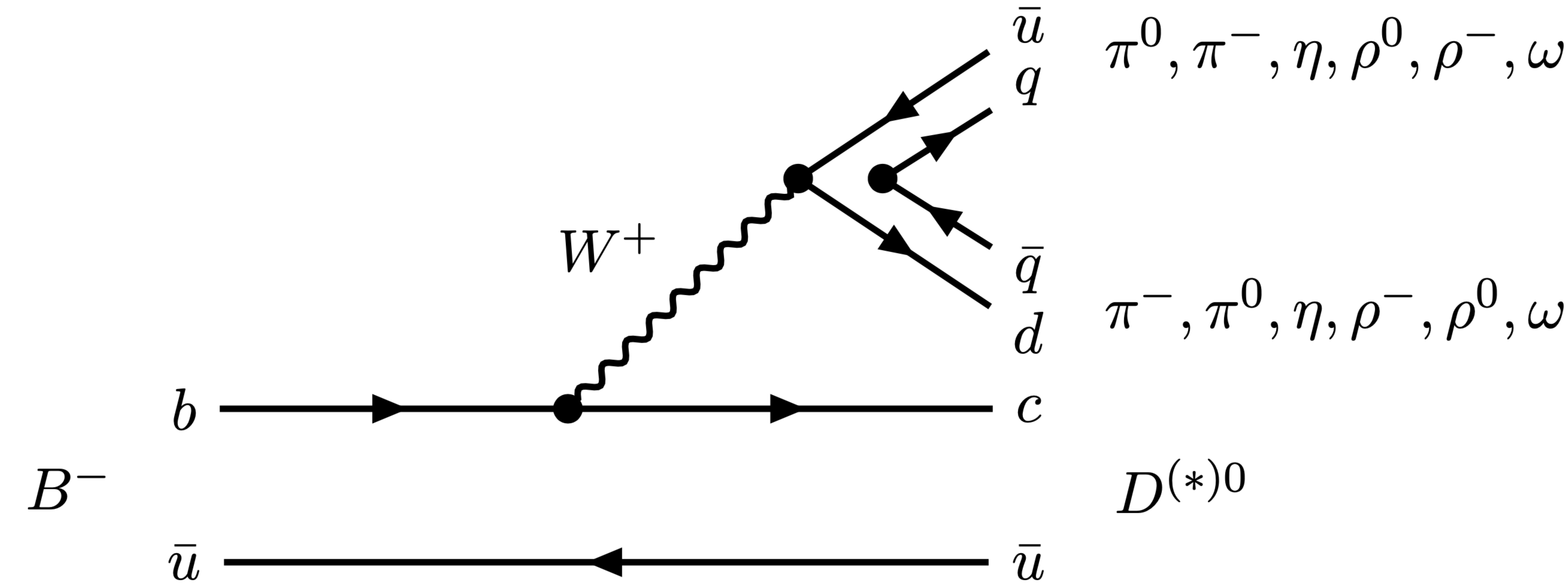}
\caption{Weak vertices obtained from the hadronization of the $\bar u d$ pair coupled to $W^-$ in the mechanism depicted in Fig.~\ref{quarklv_a}. Final states like $\pi^0\pi^- D^{*0}$, $\rho^0\pi^- D^0$, etc., can be obtained from such a mechanism.}\label{newhad}
\end{figure}

At this point, we should mention that although the hadronization of the $\bar u d$ pair coupled to $W^-$ in Fig.~\ref{quarklv_a} cannot directly produce $D^{*+}\pi^-\pi^-$ in the final state, we can obtain, among others, $D^{*0}\pi^0\pi^-$, $D^0\rho^0\pi^-$, which can, through the transitions $D^{*0}\pi^0\to D^{*+}\pi^-$, $D^0\rho^0\to D^{*+}\pi^-$, generate the final state $D^{*+}\pi^-\pi^-$ (see Fig.~\ref{newhad}). Such mechanisms could be relevant for the rescattering process described in Sec.~\ref{AR} and we have included them. Continuing with this argument, the $\pi^-$ coupled to $W^-$ in Fig.~\ref{quarklv_a} could be very well a $\rho^-$. In such a case, once more, the hadronization of the $c\bar u$ pair can give rise to $D^+\pi^-$ and the final state $D^{*+}\pi^-\pi^-$ can be obtained through the rescattering process $D^+\rho^-\to D^{*+}\pi^-$. This contribution needs to be taken into account when determining the amplitude for the $B^-\to D^{*+}\pi^-\pi^-$ process (see Sec.~\ref{AR}).
It should also be noticed that the hadronization of the $\bar c s$ pair coupled to $W^-$ in Fig.~\ref{quarklv_b} can produce final states like $D^-_s\phi D^0$ or $D^-_s\eta D^{*0}$, which through the rescattering processes $D^0\phi\to D^{*+}\pi^-$, $D^{*0}\eta\to D^{*+}\pi^-$, can give contributions to the amplitude describing the reaction $B^-\to D^-_s D^{*+}\pi^-$. However, in the context of the generation of $D_1(2420)$ via these rescattering contributions, the former are small and can be ignored, as we do.

\begin{figure}[h!]
\centering
\includegraphics[width=0.49\textwidth]{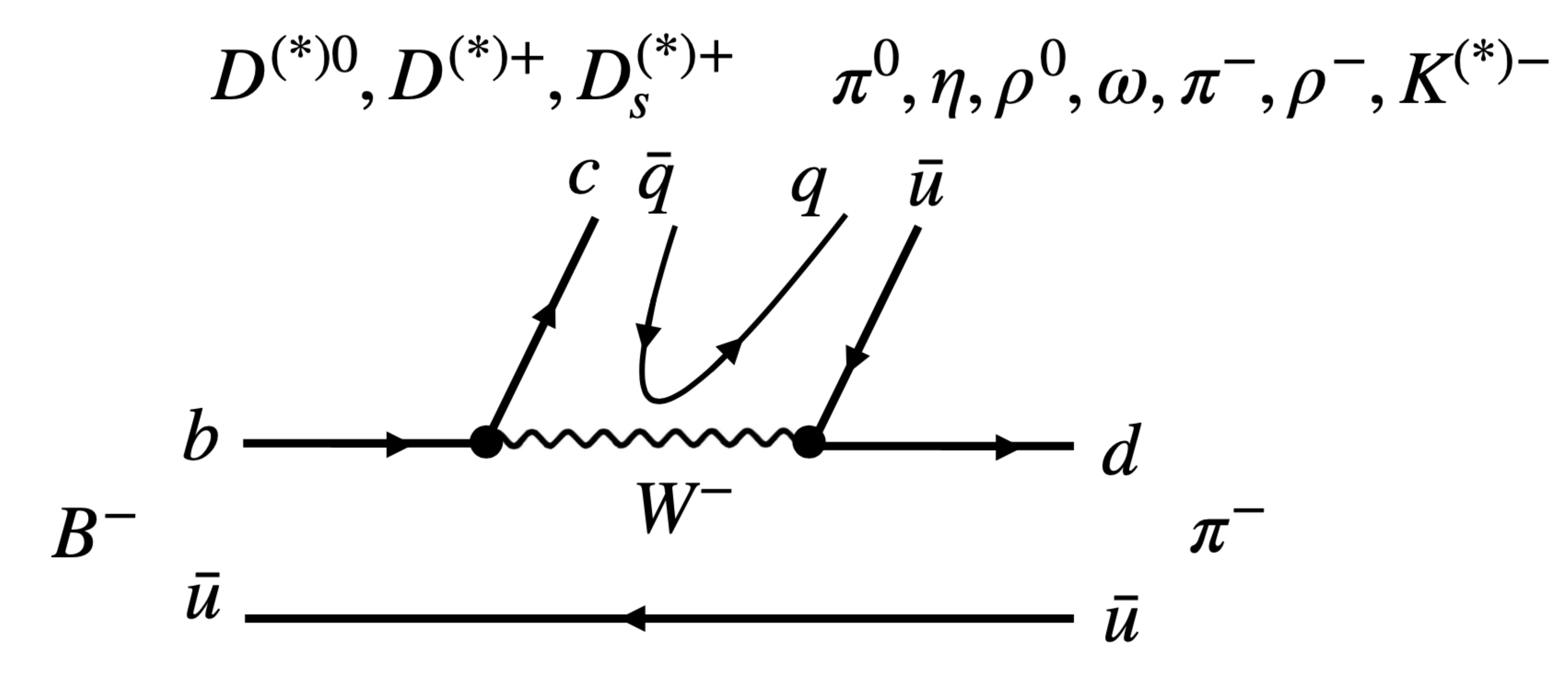}
\includegraphics[width=0.49\textwidth]{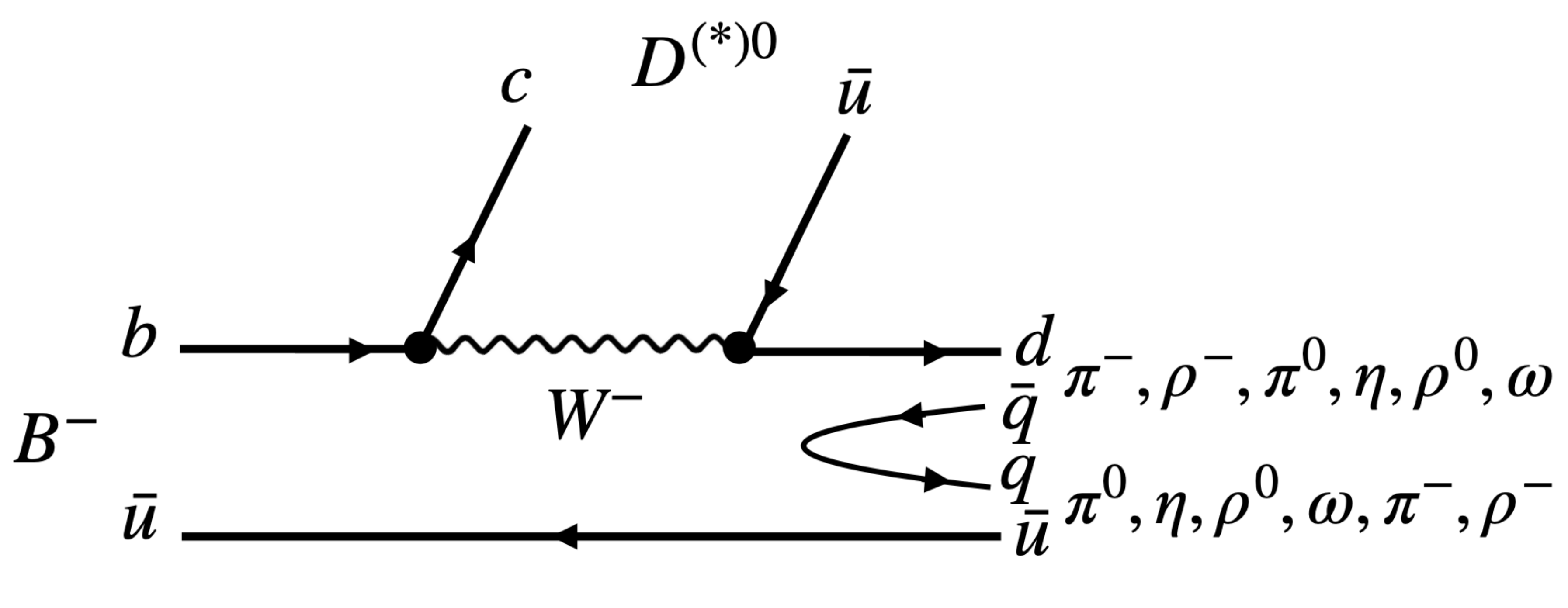}
\caption{Internal emission contributions to $B^-\to D^{*+}\pi^-\pi^-$ via tree-level (left diagram for the case in which a $D^{*+}\pi^-$ is produced from the hadronization of the $c\bar u$ pair) and rescattering (left and right diagrams giving rise to, for example, $D^{*0}\pi^0\pi^-$, $D^0\rho^0\pi^-$ final states, which, through  $D^{*0}\pi^0\to D^{*+}\pi^-$, $D^0\rho^0\to D^{*+}\pi^-$ transitions, can produce the final state $D^{*+}\pi^-\pi^-$).}\label{int}
\end{figure}

Continuing with the discussions, one peculiarity of the $B^-\to D^{*+}\pi^-\pi^-$ process when compared to $B^-\to D^-_s D^{*+}\pi^-$ is the fact that for the latter no internal emission of a $W^-$ can contribute neither at tree-level, i.e., producing directly $D^-_s D^{*+}\pi^-$, nor via rescattering, through the same hadronization procedure explained before. This is not the case of the process $B^-\to D^{*+}\pi^-\pi^-$. As can be seen in Fig.~\ref{int}, the $D^{*+}\pi^-\pi^-$ final state can be obtained from the hadronization of the $c\bar u$ pair. Such hadronization can also give rise to final states like $D^0\rho ^0\pi^-$, $D^{*0}\pi^0\pi^-$, $D^+\rho^-\pi^-$, etc., which via transitions like $D^0\rho^0\to D^{*+}\pi^-$, $D^{*0}\pi^0\to D^{*+}\pi^-$, $D^+\rho^-\to D^{*+}\pi^-$, etc., can produce the final state $D^{*+}\pi^-\pi^-$. We have also included these contributions in our calculations.
\subsection{Tree-level production amplitude}

\begin{figure}[h!]
    \centering
    \includegraphics[width=0.4\textwidth]{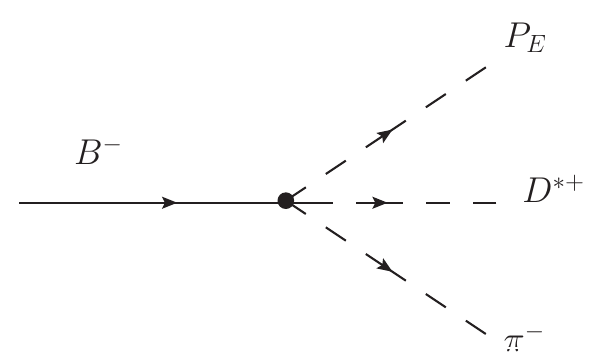}
\caption{Tree-level mechanism for $B^-\to P_E D^{*+}\pi^-$, with $P_E=\pi^-$, $D^-_s$.}
    \label{mecS_a}
\end{figure}

From the previous discussion, we can now construct the non-resonant amplitude for the $B^- \to D^{*+}\pi^{-}\pi^{-}$ and $B^{-}\to D_s^-D^{*+}\pi^{-}$ processes (see Fig.~\ref{mecS_a}). 
To determine the corresponding amplitude, we have the polarization four-vector of the $D^{*+}$ in the final state, $\epsilon^\mu_{D^*}$, which needs to be contracted with a four-vector to form a Lorentz scalar. Considering angular momentum conservation, a $P$-wave in the final state is needed to obtain total angular momentum 0 (as in the case of the $B$ meson in its rest frame) and describe the weak decays considered. This means that the mentioned polarization vector needs to be contracted with a four-momentum. In our approach, the resonance $D_1(2420)$ is obtained from $D^{*+}\pi^-$ and coupled channels interacting in the $S$-wave, and, following Refs.~\cite{Lin:2024hys,Lyu:2024zdo,Lyu:2025rsq}, we contract the mentioned polarization vector with the four-momentum of the decaying particle. In this way, we use the following amplitude to describe the contribution of the mechanism depicted in Fig.~\ref{mecS_a}:
\begin{equation}
    t^{(\text{tree})} = C_{0}p^\mu_{B} \epsilon_{D^{*+}\mu},
\label{verticeDpi}
\end{equation}
where $C_0$ is a strength parameter to be fixed, and $p^\mu_B$ is the four-momentum of the $B^-$ meson. Note that in the case where the final state is $D^{*+}\pi^-\pi^-$, the indistinguishability of the two $\pi^-$ implies a symmetrization of the tree-level amplitude, which produces a factor of two with respect to the tree-level amplitude for $B^-\to D^-_s D^{*+}\pi^-$. In the case of internal emission, the constant $C_0$ appearing in Eq.~(\ref{verticeDpi}) is considered to be $1/N_c=1/3$ ($N_c$ representing the number of colors) smaller than that for external emission~\cite{Song:2025dgg}. The relative interference (constructive/destructive) between the external and internal tree-level contributions for $B^-\to D^{*+}\pi^-\pi^-$ is fixed to obtain a better description of the experimental data. In this case, a global minus sign needs to be included in the constant associated with the tree-level contribution obtained from internal emission.

\subsection{Amplitude for the rescattering processes}\label{AR}

As discussed in the preceding subsections, the weak decay of the $B$-meson can lead via hadronization to several sets of three-meson final states of the same nature (made of mesons having the same quantum numbers). Different subsystems among these three mesons can rescatter to produce the desired final state. In particular, in the case of $B^{-}$ decaying to a final state with $D_s^-$, several vector-pseudoscalar ($VP$) systems, as listed in Eqs.~(\ref{had2}) and (\ref{had3}), can be formed at the tree-level which then could interact to generate $D^{*+}\pi^-$, alongside $D_s^-$. As discussed in Refs.~\cite{Gamermann:2007fi,Malabarba:2022pdo,Khemchandani:2023xup}, and reinforced in the introduction of this manuscript, the nonperturbative vector-pseudoscalar coupled-channel dynamics can well describe the properties  of the lightest axial meson. Hence, the inclusion of such interactions can be expected to give an important contribution when describing the related experimental data. Note that interactions among $D_s^-$ and a pseudoscalar (or vector) meson of the systems listed in Eqs.~(\ref{had2}) and (\ref{had3}) are not expected to be of much relevance since they do not have any light quark-antiquark pair in common. In the case of $B^-\to D^{*+}\pi^-\pi^-$, on the other hand, we have two sets of identical pairs where strong interactions are relevant and both should be taken into account to determine the final amplitude.

\begin{figure}[ht!]
    \centering
     \includegraphics[width=0.7\textwidth]{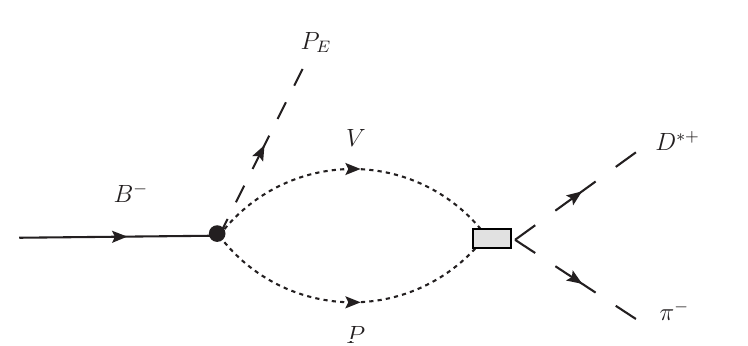}
\caption{Rescattering contribution to the $ B^-  \to D^{*+}\pi^{-}\pi^{-}$ and $B^{-}\to D_s^-D^{*+}\pi^{-}$ reactions, associated with the dynamical generation of $D_{1}(2420)$ through the vector-pseudoscalar interaction, indicated by a blob in the figure. Here, $P_{E}$ represents either $\pi^-$ or $D_s^-$ and, in the case of being $\pi^-$, the amplitude needs to be symmetrized to account for the generation of $D_1(2420)$ in any of the $D^{*+}\pi^-$ pair present in the final state.}
\label{mecS_b}
\end{figure}

In this way, the reactions considered can be produced via the rescattering mechanism shown in Fig.~\ref{mecS_b}. 

Considering the tree-level, $t^{(\text{tree})}$, and rescattering contributions discussed before, which we denote as $t^{(R)}$, the full amplitude for the processes under study can be written as 

\begin{align}
t &= t^{(\text{tree})}+t^{(R)}\nonumber\\
   &=p_B^\mu \Biggl\{(C^\text{ext}_0\omega^\text{ext}_{P_E D^{*+}\pi^-}+C^\text{int}_0\omega^\text{int}_{P_E D^{*+}\pi^-})\epsilon_{D^*\mu}- i\sum_{l=1}^8 (C^\text{ext}_0\omega^\text{ext}_l+C^\text{int}_0\omega^\text{int}_l)\nonumber\\
   &\quad\times \int\frac{d^{4}q}{(2\pi)^{4}} \left[ \sum_{\rm Pol} \epsilon_{V\mu} (q) \epsilon^{\ast}_{V\nu}(q) \right]G_{l}(q^2,m_{D^{*+}\pi^-})T_{l}(m_{D^{*+}\pi^-})\epsilon^\nu_{D^*}\Biggr\},
\label{tampS2}
\end{align}
where $m_{D^{*+}\pi^-}$ is the invariant mass of the $D^{*+}\pi^{-}$ system~\footnote{In the case of $B^-\to D^{*+}\pi^-\pi^-$, the formation of $D_1(2420)$ in both $D^{*+}\pi^-$ systems has been considered. This, not only produces a factor of two for the weight related to the $D^{*+}\pi^-\pi^-$ intermediate state, as mentioned earlier, but also $G_l(q^2,m_{D^{*+}\pi^-})T_l(m_{D^{*+}\pi^-})\to G_l(q^2,m_{D^{*+}\pi^-_1})T_l(m_{D^{*+}\pi^-_1})+G_l(q^2,m_{D^{*+}\pi^-_2})T_l(m_{D^{*+}\pi^-_2})$.}, $\omega_l$ is the weight associated with the $l$-th intermediate state produced via hadronization of the different $q_i\bar q_j$ pairs mentioned in Sec.~\ref{subweak}, and its value is (mainly) related to the coefficients for the different $VP$ systems given in Eqs.~(\ref{had2}) and (\ref{had3}). Here, the superscript ``ext'' or ``in'' in $\omega_l$ refers to the external/internal emission processes (in the case of having no contribution from internal emission processes, $C^\text{int}_0$ and $\omega^\text{int}$ are set to zero). In Table~\ref{Tw} we show the results obtained for the weights associated with $D^-_s H_1 H_2$ (here $H_i$ represents a hadron) and $H_1 H_2\pi^-$ intermediate states. For the latter, the results shown in Table~\ref{Tw} consider the hadronization of the $c\bar u$ pair in Fig.~\ref{quarklv_a} and of that of the $\bar u d$ pair in Fig.~\ref{newhad}. Also, in the case of $B^-\to D^{*+}\pi^-\pi^-$,  the weights obtained for the external and internal emission processes coincide in modulus. At this point, we should mention that the phase related to the interference of the vector-pseudoscalar channels present in Eq.~(\ref{had2}) with those of Eq. (\ref{had3}) cannot be obtained from the procedure followed and we need to fix it when reproducing the data. Considering the two most common possible outcomes, i.e., constructive/destructive interference between these channels, we find that, for the external emission processes, a global minus sign (i.e., destructive interference) in the $D^-_s P_H V_L$ weights [here $P_{H/L}$ ($V_{H/L}$) means heavy/light pseudoscalar (vector) meson] listed in Table~\ref{Tw} is needed. In the case of the internal emission processes, this global minus sign needs to be considered in the $P_L V_H\pi^-$ weights shown in Table~\ref{Tw}. We should also mention here that the hadronization procedure producing $D^+\rho^-\pi^-$, $D^{*0}\pi^0\pi^-$, and $D^{*0}\eta\pi^-$ has contributions not only due to the formation of the vector and one of the pseudoscalars from the hadronization, but also from the generation of the two pseudoscalars as a consequence of the hadronization. In the former case, the corresponding weights are obtained from the product of the matrices $P\cdot V$ or $V\cdot P$, as explained earlier, and in the latter case, the product of the matrices $P\cdot P$ is needed. To this respect, we have followed Ref.~\cite{Song:2025dgg} to calculate these weights, where the $W\to P_1P_2$ vertex, with $P_1$, $P_2$ representing two pseudoscalar particles, is given by a Lagrangian $\mathcal{L}\sim W_\mu (\bar P_1\partial_\mu \bar P_2-\partial_\mu\bar P_1\bar P_2)$. In this way, we find only contribution to the $D^+\rho^-\pi^-$ channel from the hadronization mechanism giving rise to the two pseudoscalars present.

\begin{table}
\caption{Weights related to the intermediate states contributing to the weak processes considered}\label{Tw}
\begin{tabular}{cccc}
\text{channel}&\text{weight}&\text{channel}&\text{weight}\\
$D^-_s D^{*+}\pi^-$& 1& $D^{*0}\pi^0\pi^-$& $1/\sqrt{2}$\\
$D^-_s D^{*0}\eta $& $1/\sqrt{3}$& $D^{*0}\eta\pi^-$& $1/\sqrt{3}$\\
$D^-_s D^{*+}\pi^-$& 1& $D^{*+}\pi^-\pi^-$ &2\\
$D^-_s D^{*+}_s K^-$& 1& $D^{*+}_s K^-\pi^-$& 1\\
$D^-_s D^{0} \rho^0$& $1/\sqrt{2}$& $D^{0}\rho^0\pi^- $ & $1/\sqrt{2}$\\
$D^-_s D^{0} \omega$& $1/\sqrt{2}$&$D^{0} \omega\pi^-$ & $3/\sqrt{2}$\\
$D^-_s D^{+} \rho^-$& 1& $D^+\rho^-\pi^-$& 2\\
$D^-_s D^{+}_s K^{*-}$& 1& $D^+_s K^{*-}\pi^-$& 1\\
\end{tabular}    
\end{table}

It is useful to mention here that we use the approximation   
$T_{l}(m_{D^{*+}\pi^-})\epsilon^{\ast}_{V\nu}\cdot\epsilon^\nu_{D^*} \simeq -T_{l}(m_{D^{*+}\pi^-}){\epsilon}^{\ast}_{Vi}\cdot{\epsilon}_{D^{*}}^i$, where the minus sign comes from the non-relativistic limit, in which we consider ${\epsilon}^{\ast}_{V_0}\cdot{\epsilon}^{0}_{D^{*}} << 1$. In Eq.~(\ref{tampS2}),
\begin{align}
G_{l}(q^2,m_{D^{*+}\pi^-})=\frac{1}{q^2-m^2_{V_l}+i\epsilon}\frac{1}{(p_B-p_E-q)^2-m^2_{P_l}+i\epsilon} 
\end{align}
is the product of the propagators for the vector and pseudoscalar mesons in the intermediate state, with $q$ being the four-momentum of the vector meson of mass $m_{V_l}$ and whose polarization vector is $\epsilon_V$. The symbol $\sum\limits_{\rm Pol}$ in Eq.~(\ref{tampS2}) denotes the sum over the polarizations of the intermediate vector meson; $T_{l}$ represents the $V_l P_l \to D^{*+}\pi^{-}$ element of the unitarized transition matrix obtained by solving the Bethe-Salpeter equation:
\begin{equation}
    T = \left[1 - VG\right]^{-1}V,
\label{eq8}
\end{equation}
where $V$ is a matrix whose elements represent the $S$-wave $VP\to VP$ amplitudes obtained from an effective Lagrangian based on the relevant symmetries of the problem under study. In Eq.~(\ref{eq8}), $G$ is a diagonal matrix whose elements represent the two-hadron loop functions for the coupled channels considered when solving the equation. The details of the calculation of $T$ can be found in Refs.~\cite{Malabarba:2020grf,Khemchandani:2023xup}. 
It is important to remind the reader here that, as discussed in the Introduction, the Bethe-Salpeter equation was solved in Ref.~\cite{Khemchandani:2023xup} using meson-meson contact interactions, box diagrams, and a bare quark-model pole. The latter was added as an attempt to improve the agreement between a wide state found in the model with the mass and width of the $D_1(2430)$ listed in Ref.~\cite{ParticleDataGroup:2024cfk}.  However, our present work aims to reproduce the lineshape of $D_1(2420)$ in the data, that gets well described in terms of two-meson dynamics. Besides, we would like to keep the number of parameters to be as small as possible, especially keeping in mind that many tensor integrals are to be regularized in the formalism (see Appendix~\ref{app:amplbroad} for more details). Hence, we drop the quark model pole here, since we do not aim at determining the properties of $D_1(2430)$ and we can then eliminate the corresponding unknown parameters related to it. 

The final form of  $ t^{(R)}(m_{D^{*+}\pi^-})$ in Eq.~(\ref{tampS2}), needed to determine the \textbf{$D^*\pi$} mass distribution, is obtained through a series of mathematical manipulations detailed in Appendix~\ref{app:amplbroad}.

Once the amplitude $t$ describing the weak decay $B^-\to D^{*+}\pi^- P_E$ is determined, the $D^{*+}\pi^-$ invariant mass distribution can be obtained from the following expression~\cite{ParticleDataGroup:2024cfk}:
\begin{equation}
    \frac{d\Gamma}{d m_{12}} = \frac{1}{(2\pi)^3}\frac{m_{12}}{8 m^{3}_{B}}\ \int\limits_{\left(m_{23}\right)_{\text{min}}}^{\left(m_{23}\right)_{\text{max}}} dm_{23} \  m_{23}  \sum_{\text{Pol}}\left\vert t \right\vert ^{2}  ,
\label{Smassdistri}
\end{equation}
where $t$ is given by Eq.~(\ref{tampS2}), and the symbol $\sum\limits_{\text{Pol}}$ indicates summation over the polarizations of the $D^{*+}$ meson in the final state. In Eq.~\eqref{Smassdistri}, the limits $\left(m_{23}\right)_{\text{min}}$ and $\left(m_{23}\right)_{\text{max}}$ are given by
\begin{eqnarray}
    (m_{23})_{\textrm{min}} & = & \sqrt{(E^{*}_{2} + E^{*}_{3})^2 - \bigg(\sqrt{E^{*\,2}_{2} - m_{2}^{2}} + \sqrt{E^{*\,2}_{3} - m_{3}^{2}} \bigg)^2}, \nonumber \\    
    (m_{23})_{\textrm{max}} & = & \sqrt{(E^{*}_{2} + E^{*}_{3})^2 - \bigg(\sqrt{E^{*\,2}_{2} - m_{2}^{2}} - \sqrt{E^{*\,2}_{3} - m_{3}^{2}} \bigg)^2},\label{M_lim}
\end{eqnarray}
with
\begin{eqnarray}
    E^{*}_{2} = \frac{m_{12}^{2} - m_{1}^{2} + m_{2}^{2}}{2m_{12}}, \nonumber \\    
    E^{*}_{3} = \frac{m_{B}^{2} - m_{12}^{2}- m_{3}^{2}}{2m_{12}}.\label{E_star}
\end{eqnarray}
In Eq.~\eqref{E_star}, $E^{*}_{2}$ and $E^{*}_{3}$ are the energies of particles $2$ and $3$ evaluated at the center-of-mass frame of the subsystem $12$. For the process with $D_s^-$ in the final state, we label $
1\to D^{*+}$, $2\to\pi^{-}$ and $3\to D_s^{-}$. In the case of having two pions in the final state, we define $1\to\pi^{-}$, $2\to D^{*+}$ and $3\to\pi^{-}$, and symmetrize the amplitude~\footnote{In this case, $d\Gamma/dm_{D^{*+}\pi^-}$ is given by the sum $d\Gamma/dm_{12}+d\Gamma/dm_{23}$. Normalization factors due to identical particle have been reabsorbed in the value of $C^\text{ext}_0$ appearing in the amplitude. Also, as explained in Sec.~\ref{Res}, the fact that in the experiment it is the invariant mass distribution $d\Gamma/dm(D^{*+}\pi^-)_\text{low}$ the one which shows the signal for $D_1(2420)$ makes that a Heaviside-theta function $\theta(m_{23}-m_{12})$ needs to be included when integrating on $m_{23}$ in Eq.~(\ref{Smassdistri}) to guaranty that $m_{23}> m_{12}$. Similarly, for calculating $d\Gamma/dm_{23}$, a Heaviside-theta function $\theta(m_{12}-m_{23})$ needs to be considered when integrating on $m_{12}$.}. In this way, using Eq.~\eqref{Smassdistri} we calculate the $D^{*+}\pi^-$ invariant mass distribution for both weak decays.

\section{Results and discussions}\label{Res}
\begin{figure}[h!]
\centering
\includegraphics[width=0.45\textwidth]{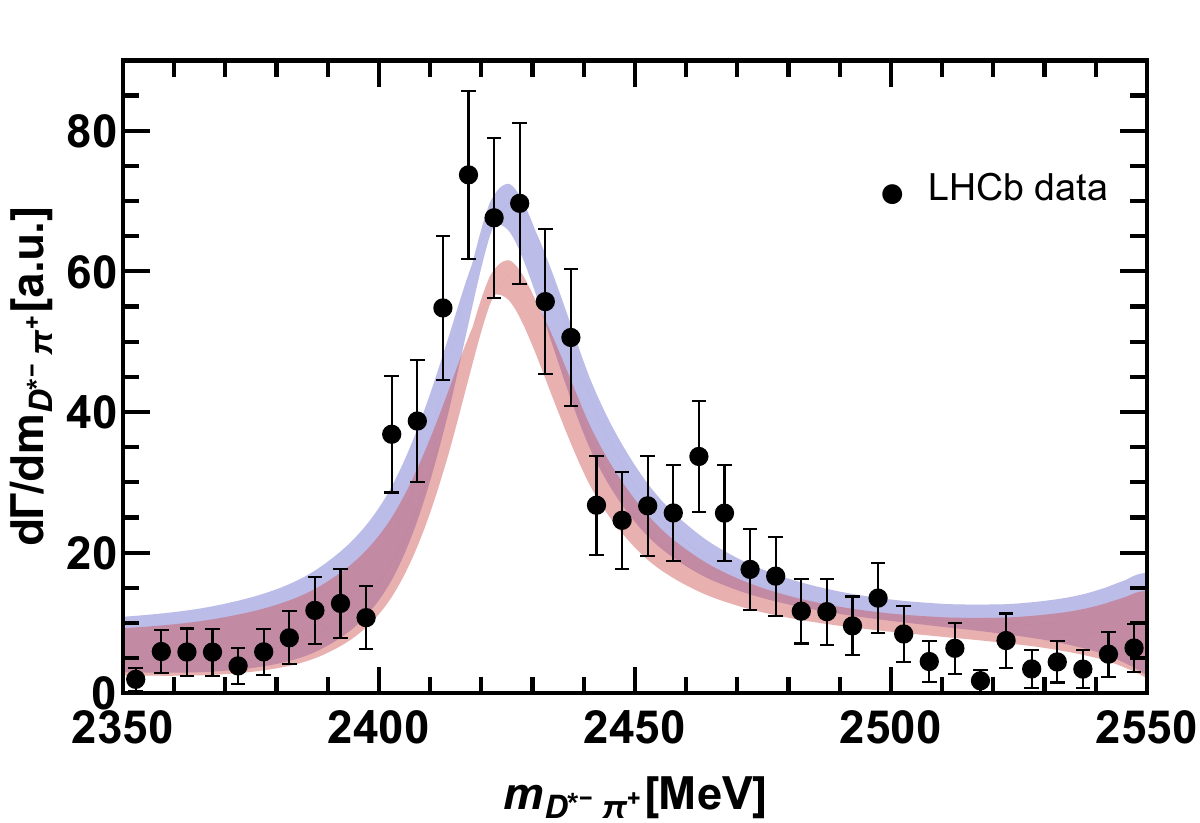}
\includegraphics[width=0.45\textwidth]{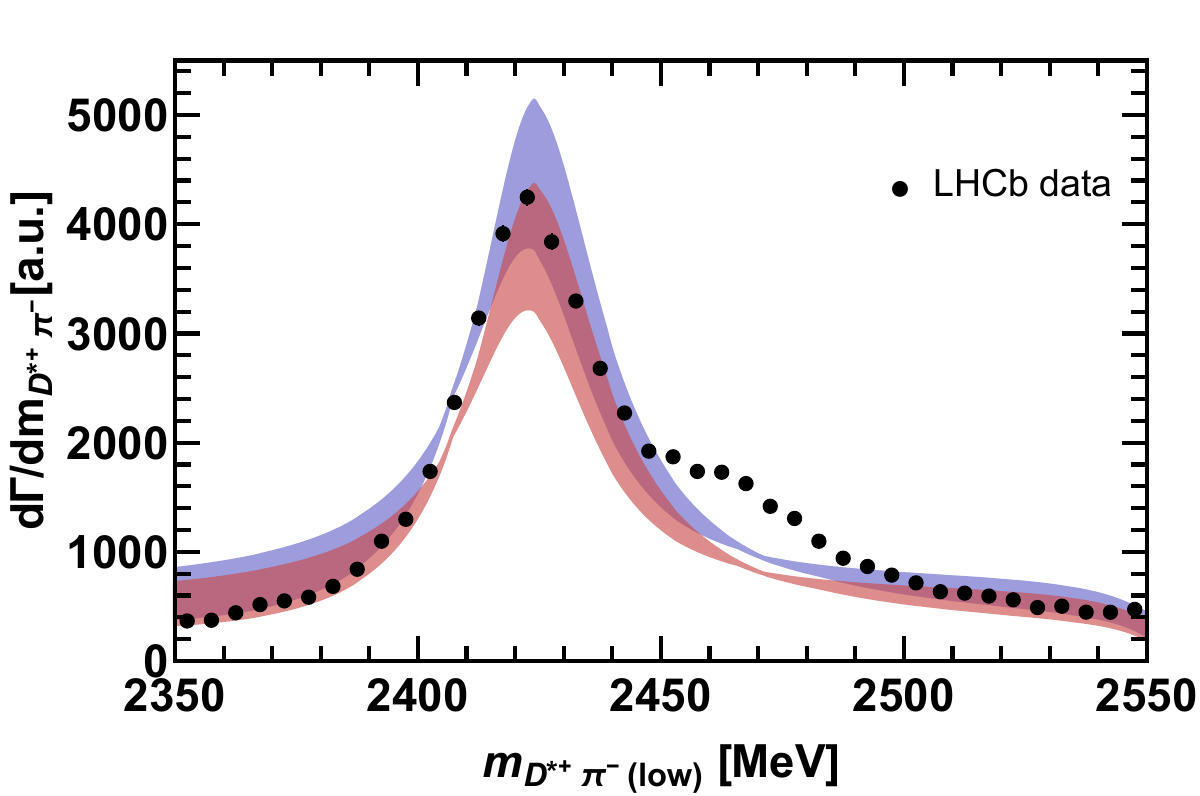}
\caption{Invariant mass distribution for $D^{*+}\pi^-$ in the process $B^-\to D^-_s D^{*+}\pi^-$ (left) and $B^-\to D^{*+}\pi^-\pi^-$ (right). The experimental data are taken from Refs.~\cite{LHCb:2024vhs} (left panel) and ~\cite{LHCb:2019juy} (right panel). The bands shown represent the uncertainties related to the theoretical model considering different form factors to regularize the integrals appearing in the calculation of the amplitude $t$ involved in Eq.~(\ref{Smassdistri}) (we refer the reader to the Appendix~\ref{app:amplbroad} for more details). The same form factors have been used in both figures. The higher (lower) band is obtained considering a normalization of the theoretical curve to $100\,(85)\%$ of the total area of the experimental (central points) distribution (more details are given in the text). The meaning of $m_{D^{*+}\pi^-(\text{low})}$ can be found in the main text in the discussions related to this figure.}\label{DistriDstarpi}
\end{figure}

In Fig.~\ref{DistriDstarpi} (left) we show the result obtained for the $D^{*+}\pi^-$ invariant mass distribution in the process $B^-\to D^-_s D^{*+}\pi^-$. As mentioned in Sec.~\ref{sec:level2}, no internal emission mechanisms of $W^-$ contribute to this process. The constant $C^\text{ext}_0$ present in Eq.~(\ref{Smassdistri}), through Eq.~(\ref{tampS2}), has been fixed to reproduce the area under the experimental curve, which is obtained by considering the central values of the data. This way of fixing this constant assumes that the width for the weak decay processes considered is only due to the mechanisms explained in Sec.~\ref{sec:level2}, in which $D_1(2420)$ is generated in the $S$-wave via rescattering. Although the signal observed for $D_1(2420)$ in the $D^{*+}\pi^-$ invariant mass distribution of the processes $B^-\to P_E D^{*+}\pi^-$ ($P_E=D^-_s$, $\pi^-$) dominates in the spectrum, the hypothesis considered ignores contributions to the partial decay width of $B^-\to P_E D^{*+}\pi^-$ from other resonances different to $D_1(2420)$. In view of this, we have also taken into account the possibility of the $S$-wave signal obtained for $D_1(2420)$ from the rescattering of vector-pseudoscalars to be responsible for $85\%$ of the width related to the process $B^-\to P_E D^{*+}\pi^-$. The bands shown in Fig.~\ref{DistriDstarpi} correspond to the  uncertainties in the model arising from different choices of the form factors used to regularize the integrals involved in the calculation of Eq.~(\ref{tampS2}) (details on technical calculations are given in the Appendix~\ref{app:amplbroad}). The higher band is obtained by fixing $C^\text{ext}_0$ to the total area of the distribution, while the lower band represents $85\%$ of this area. We should state here that the way of fixing the constant $C^\text{ext}_0$ produces a mild dependence of its value with the cut-off $\Lambda$ and type of form factor used to regularize the integrals (see Table~\ref{Cval}). As can be seen, the mechanism depicted in Fig.~\ref{mecS_b}, in which $D_1(2420)$ is generated from $S$-wave vector-pseudoscalar interactions, with its subsequently decay in the $S$-wave to $D^{*+}\pi^-$, is responsible, within uncertainties, for a relevant part of the signal observed in the spectrum for the mentioned $D_1$ resonance. 

\begin{table}[h!]
\centering
\caption{Some of the values found for $C^\text{ext}_0$ for different form factors and $\Lambda$'s when considering $100\%$ of the area of the experimental distribution. The first (second) column of values under the $C^\text{ext}_0$ entry is related to $B^-\to D^-_s D^{*+}\pi^-$ ($B^-\to D^{*+}\pi^-\pi^-$)}\label{Cval}
\begin{tabular}{c|c|cc}
Form Factor& $\Lambda$ (MeV)& \multicolumn{2}{c}
{$C^\text{ext}_0$ (MeV${}^{-1}$)}\\\hline
\multirow{2}{*}{Gaussian}&$900$&$0.27$&$2.58$\\
&$1100$&$0.19$&$2.07$\\
\multirow{2}{*}{Monopole}&$600$&$0.21$&$2.32$\\
&$900$&$0.13$&$1.70$
\end{tabular}
\end{table}
Similar conclusions can be drawn for the $D^{*+}\pi^-$ invariant mass distribution in the process $B^-\to D^{*+}\pi^-\pi^-$, shown in Fig.~\ref{DistriDstarpi} (right). In this case, as a consequence of the indistinguishable negative charged pions in the final state, there are two $D^{*+}\pi^-$ invariant masses corresponding to the identical pairs $12$ and $23$, which are symmetric under the interchange $1\leftrightarrow 3$. In such a scenario, as stated in Ref.~\cite{LHCb:2019juy}, two invariant masses are to be defined, $m_{D^{*+}\pi^-(\text{low})}$ and $m_{D^{*+}\pi^-(\text{high})}$, which describes the situation that when one of the identical pairs has lower invariant mass the other pair has higher value of the invariant mass. Discussions on another process with identical particles, requiring similar definitions of low/high invariant masses can be found in Refs.~\cite{Wang:2022xqc,Song:2025dgg,LHCb:2021enr,BESIII:2025yag}.
In Ref.~\cite{LHCb:2019juy}, the prominent presence of the resonances $D_1(2420)$ and $D^*_2(2460)$ were observed in the $m_{D^{*+}\pi^-(\text{low})}$ invariant mass distribution; thus, we consider the latter one here.  As can be seen in Fig.~\ref{DistriDstarpi} (right), a $S$-wave $D_1(2420)\to D^{*+}\pi^-$ process, in which $D_1(2420)$ is generated from the rescattering mechanism shown in Fig.~\ref{mecS_b}, is responsible for an important part of the signal observed for $D_1(2420)$ in the $m_{D^{*+}\pi^-(\text{low})}$ invariant mass distribution. In this case, as explained in Sec.~\ref{sec:level2}, both external and internal emission processes contribute to this invariant mass distribution. As mentioned before, we consider $C^\text{int}_0=-C^\text{ext}_0/3$, with $C^\text{ext}_0$ being fixed either to reproduce the total area under the experimental distribution (central points) or $85\%$ of its value. The upper (lower) band in Fig.~\ref{DistriDstarpi} (right) represents the first (second) choice, i.e., $100\%$ ($85\%$) of the mentioned area. We should mentioned here that the consideration of the internal emission processes helps in finding a result more in line with the experimental data, especially at lower values of the invariant mass distribution. In this context, it should be noted that the normalization procedure followed does not alter the resonance signal/background proportion of the theoretical results for either of the two weak processes investigated, and we could have very well adjust $C^\text{ext}_0$ to reproduce the magnitude of the invariant mass distribution at a certain value of $m_{D^{+*}\pi^-}$.

\begin{figure}[h!]
\centering
\includegraphics[width=0.49\textwidth]{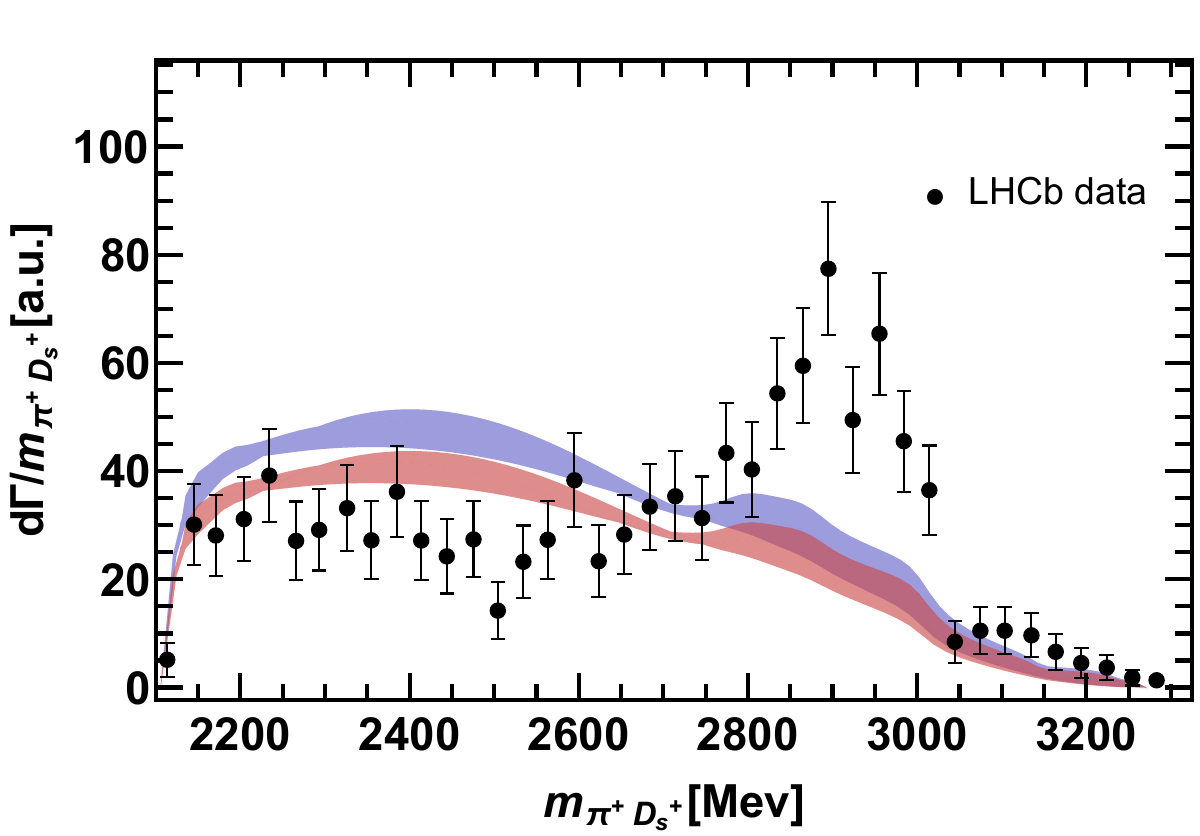}
\includegraphics[width=0.49\textwidth]{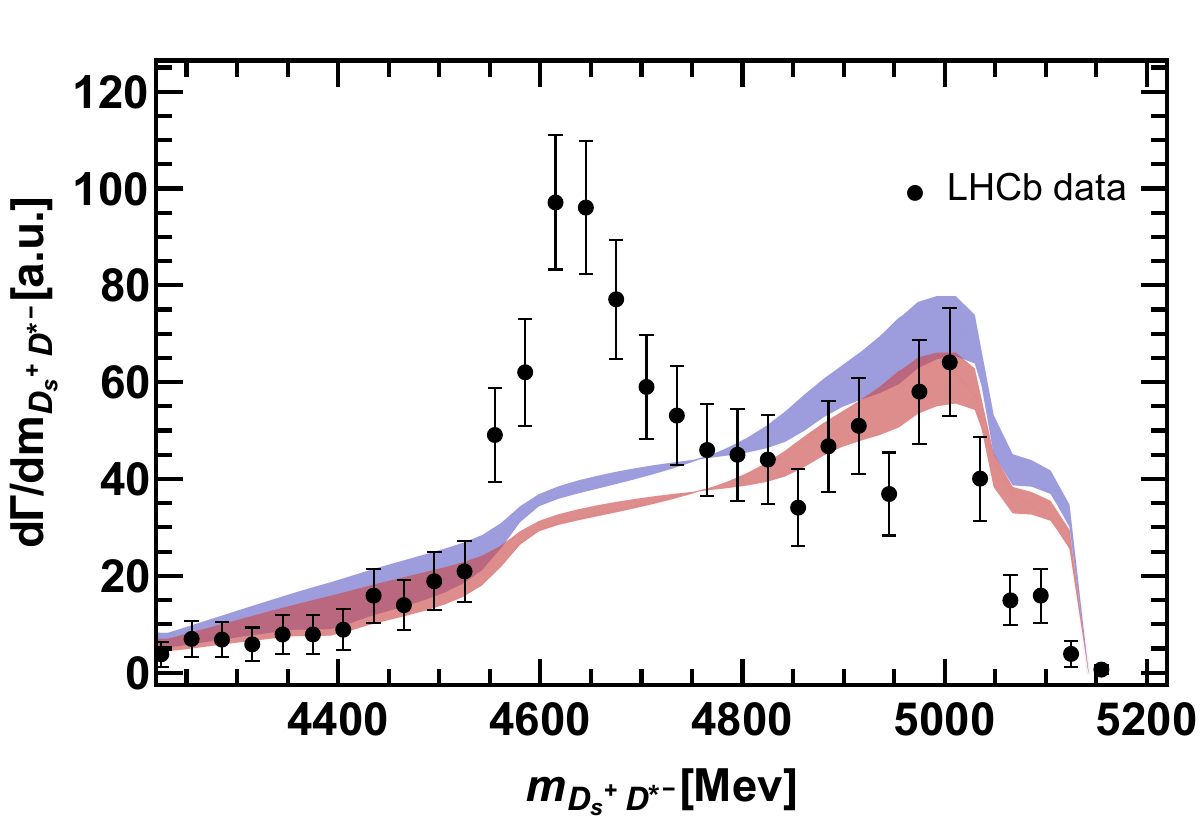}
\includegraphics[width=0.5\textwidth]{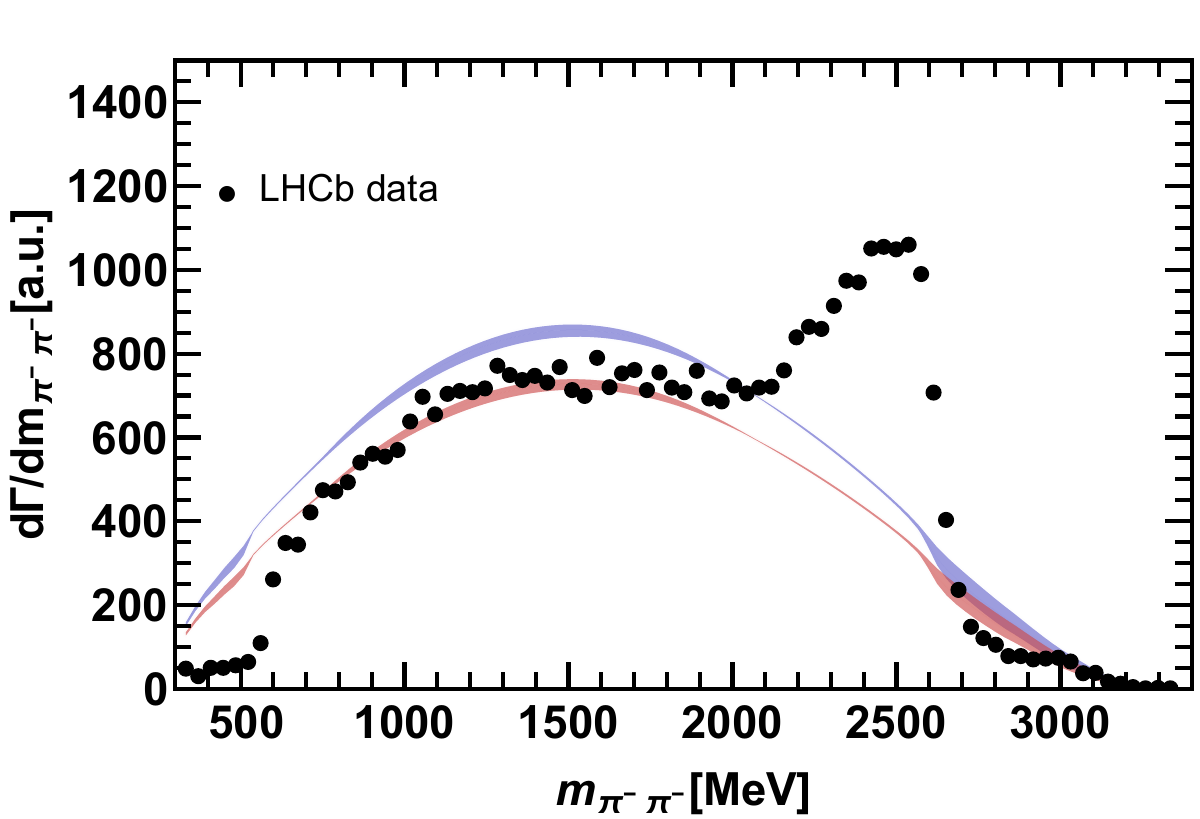}
\caption{Other invariant mass distributions. The higher and lower bands have the same meaning as those in Fig.~\ref{DistriDstarpi}. Although we consider the charge conjugated reaction of $B^+\to D^+_s D^{*-}\pi^+$, no evident CP violation effects are observed in Ref.~\cite{LHCb:2024vhs}. Thus, we can consider that the first two distributions shown and their respective charge conjugated version are equivalent.}\label{others}
\end{figure}

With the value of $C^{\text{ext}}_0$ fixed for each of the weak processes studied, we can determine the $D^-_s\pi^-$, $D^-_s D^{*+}$ and $\pi^-\pi^-$ invariant mass distributions of the processes $B^-\to D^-_s D^{*+}\pi^-$ and $B^-\to D^{*+}\pi^-\pi^-$. The results are shown in Fig.~\ref{others}. As can be seen, with the simple model considered, we can reproduce reasonably well the bulk of the data, although the agreement with the experimental data is worse when compared to that of Fig.~\ref{DistriDstarpi}. This can be due to the simplicity of the model considered, where (1) the explicit interaction of the pair of particles constituting these respective invariant masses have not been taken into account. This, although it might not affect much the $D^{*+}\pi^-$ invariant mass distribution, could have some influence in the invariant mass distributions of the others pairs; and (2) these invariant mass distributions could be more sensitive to the presence of other resonances different to $D_1(2420)$. A better description of the invariant mass distributions studied here could be obtained by including the mentioned interactions between the corresponding pair of particles forming the respective invariant masses and/or including other $D_1$, $D^*_2$, $D_0$, $D^*_1$, $D_2$ and $D^*_3$ resonances which have masses in the range $\sim 2400-2800$ MeV. A constructive/destructive interference of such additional contributions with the amplitudes considered here can contribute to the invariant mass distributions with different partial waves, as done in the amplitude analyses of Refs.~\cite{LHCb:2024vhs,LHCb:2019juy}. This is beyond the scope of this work, and it is also not our purpose, which is to investigate whether a $D_1(2420)$ dynamically generated from $S$-wave vector-pseudoscalar  interactions can be responsible, or not, for the signal observed in the $D^{*+}\pi^-$ invariant mass distribution of the processes $B^-\to D^-_s D^{*+}\pi^-$, $B^-\to D^{*+} \pi^-\pi^-$. In this context, we should clarify that we have not tried to make a fit of the parameters present in the model to the experimental data to minimize, for example, the $\chi^2$ per number of degrees of freedom (n.d.o.f): we have simply make some choices, like fixing $C^{\text{ext}}_0$ to the area of the distribution, choosing a constructive/destructive interference  between $D^-_s P_H V_L$ and $D^-_s P_L V_H$ intermediate states contributing to $B^-\to D^-_s D^{*+}\pi^-$ via external emission and between $P_H V_L\pi^-$, $P_L V_H\pi^-$ for the external and internal emission processes contributing to $B^-\to D^{*+} \pi^-\pi^-$. We have not even tried to consider more general phases of the type $e^{i\delta_j}$ and fix $\delta_j$ by minimizing the $\chi^2/\text{n.d.o.f}$. Same thing happens for the cut-off present in the form factors used to regularize the loops. We have simply varied it in a reasonable range considering the fact that $D_1(2420)$ can be interpreted as a kind of molecular state generated from vector-pseudoscalar dynamics. Thus, the overall agreement obtained with the experimental data is not trivial. For instance, using the highest curve composing the higher band in Fig.~\ref{DistriDstarpi} (left), the calculation of $\chi^2/\text{n.d.o.f}$, considering the theoretical result obtained as if it came from a fitting of the parameters to the data, gives a value of $\sim 2$ restricting $m_{D^{*+}\pi^-}\sim [2355-2500]$ MeV. Such a result suggests that fixing the parameters by a $\chi^2$ minimization procedure could probably produce a better value of $\chi^2/\text{n.do.f}$. Regardless of this, our calculations show that the generation of $D_1(2420)$ through  $S$-wave interactions of vectors and pseudoscalars and its $S$-wave decay to $D^*\pi$ provides a significant contribution. Such a conclusion gives a different perspective on the $D_1(2420)$ decay dynamics than that of Refs.~\cite{LHCb:2019juy,LHCb:2024vhs}. The analyses carried out in Refs.~\cite{LHCb:2019juy,LHCb:2024vhs} reported a dominant $D^*\pi$ contribution coming through a decay in  $D$-wave, when compared to the  decay in the $S$-wave within their multi-resonance fits ($\sim 56-80 \%$ versus $1.2-7.6\%$ in Ref.~\cite{LHCb:2024vhs} and $\sim 57-63\%$ versus $2.5-9.3\%$ in Ref.~\cite{LHCb:2019juy}). The  mechanism in our study, generating $D_1(2420)$ from $S$-wave vector-pseudoscalar dynamics, suggests that a potentially more significant contribution to the $D^*\pi$ invariant mass distributions could come from the decay of $D_1(2420)$ in the $S$-wave, offering a complementary viewpoint that may help disentangle the partial-wave structure of the signal observed in future experimental analyses.

Next, we can determine the isospin $1/2$ $D^*\pi$ scattering-length from the two-body $VP\to D^{*+}\pi^-$ $T$-matrix used to describe the rescattering mechanism illustrated in Fig.~\ref{mecS_b}. We write the $D^*\pi $ state in the isospin $1/2$ basis as
\begin{equation}
| D^*\pi \rangle= \sqrt{\frac{2}{3}}| D^{*+}\pi^-\rangle +\frac{1}{\sqrt{3}} |D^{*0}\pi^0 \rangle,
\end{equation}
and calculate the corresponding $D^*\pi$ scattering length, $a_{D^*\pi}$, using the $D^*\pi\to D^*\pi$ $T$-matrix in isospin $1/2$ through
\begin{align}
a_{D^*\pi}=\frac{1}{8\pi m^{\text{th}}_{D^*\pi}}T_{D^*\pi\to D^*\pi}(m^{\text{th}}_{D^*\pi}).\label{aDstarpi}
\end{align}
In the previous equation,  $m^{\text{th}}_{D^*\pi}=m_D^*+m_\pi$ represents the invariant mass of the $D^*\pi$ system at the threshold. Before continuing with further discussions, we should mention here that, although the values used for the $\Lambda$ appearing in the calculation of the invariant mass distribution are in line with the corresponding values related to the subtraction constant used to regularize the $G$ in Eq.~(\ref{eq8}), the latter has been regularized independently of the integrals appearing in the calculation of the invariant mass distribution with the aim of fixing the basic properties of $D_1(2420)$, such as mass and width. By doing this, the result obtained for the mentioned scattering length is $a_{D^*\pi}=-0.45$ fm (typical uncertainties for the scattering length related to the variation of the subtraction constant appearing in the $G$ of Eq.~(\ref{eq8}) are of the order of $10\%$). The value obtained for $a_{D^*\pi}$ is in agreement with that obtained in Ref.~\cite{Liu:2012zya}, recalling that the sign convention followed in Ref.~\cite{Liu:2012zya} is opposite to the one given in Eq.~(\ref{aDstarpi}). It is useful to remind the reader here that the value on the $D^*\pi$ scattering-length determined from the experimental data on the correlation function~\cite{ALICE:2024bhk} differs from the one obtained in Ref.~\cite{Liu:2012zya} using lattice data and a chiral unitary approach.

\section{Conclusions}
In this work, we have determined the $D^{*+}\pi^-$ invariant mass distribution in the decay processes $B^-\to D^-_s D^{*+}\pi^-$, $D^{*+}\pi^-\pi^-$. Considering a tree-level mechanism to produce the particles in the final state, as well as the rescattering of other vector and pseudoscalar mesons that can be produced in the intermediate state to eventually produce $D^{*+}\pi^-$, we can describe  the $D^{*+}\pi^-$ invariant mass distributions obtained from both weak decay processes reasonably well. The mentioned rescattering mechanism generates the axial meson $D_1(2420)$ via $S$-wave interactions. The reproduction of the mentioned invariant masses shows the relevant role that the $S$-wave $D^*\pi$ and coupled channel interactions play in the decay processes investigated. 
This result offers an alternative point of view to the interpretation presented in Refs.~\cite{LHCb:2019juy,LHCb:2024vhs}. Those analyses, based on fits to the experimental data, favor a description in which $D_1(2420)$ decays into $D^{*+}\pi^-$ predominantly through a $D$-wave. However, those works neglect the contribution of other coupled channels with further rescattering, which we consider here.

We have also determined the $D^*\pi$ scattering length in the isospin 1/2, by treating the data on  the $D^{*+}\pi^-$ invariant mass distribution  as an alternative source to constrain our model amplitudes. The resulting value is found to be in agreement with the one determined from the chiral unitary approach of Ref.~\cite{Liu:2012zya}, where the model parameters are constrained by the lattice data.

\begin{acknowledgements}

We are indebted to Eulogio Oset and Rafael Silva Coutinho for very useful discussions. This work was partly supported by the Brazilian agencies CNPq  (Conselho Nacional de Desenvolvimento Cient\'ifico e Tecnol\'ogico) (K.P.K: Grants No. 306461/2023-4; A.M.T: Grant No. 304510/2023-8; L.M.A.: Grants No. 400215/2022-5, 308299/2023-0, 402942/2024-8; B.A: Grant No.
987654/2023-1, 200204/2025-4), Fapesp (Funda\c c\~ao de Amparo \`a Pesquisa do Estado de S\~ao Paulo) (A.M.T. Grant No. 2023/01182-7), and CNPq/FAPERJ under the Project INCT-F\'isica Nuclear e Aplica\c{c}\~oes (Contracts No. 464898/2014-5 and 408419/2024-5). P.B  gratefully acknowledges the partial support provided by the Coordena\c{c}\~{a}o de
Aperfei\c{c}oamento de Pessoal de N\'{\i}vel Superior -- Brasil
(CAPES) -- Finance Code 001.

\end{acknowledgements}

\appendix
\section{Calculation of $t^{(R)}$}\label{app:amplbroad}

Here we summarize the main aspects of the evaluation of the amplitude $t^{(R)}$ in Eq.~(\ref{tampS2}). Using Eq.~(\ref{verticeDpi})  and performing the sum over the polarizations of the intermediate vector meson in the diagram of Fig.~\ref{mecS_b}, the rescattering contribution in Eq.~(\ref{tampS2}) can be written as   
\begin{eqnarray}
t^{(R)} = -i \sum\limits_{l = 1}^8 (C^\text{ext}_0\omega^\text{ext}_l+C^\text{int}_{0}\omega^\text{int}_l) I_{l\nu}T_{l}(m_{D^{*+}\pi^-})\epsilon^\nu_{D^{*+}},
\label{amplB1}
\end{eqnarray}
where $I_{l\nu}$ is given by
\begin{eqnarray}
    I_{l\nu} & \equiv & \int\limits^{+\infty}_{-\infty} \frac{d^{4}q}{(2\pi)^{4}}p_{B}^{\mu}\Bigg(-g_{\mu\nu} + \frac{q_{\mu}q_{\nu}}{m^{2}_{V_{l}}}\Bigg)\frac{1}{q^{2} - m^{2}_{V_{l}} + i\epsilon}\frac{1}{(p_{B} - p_{E}-q)^{2} - m^{2}_{P_{l}} + i\epsilon} \nonumber \\
    & \equiv & - I^{(a)}_{l\nu} +\frac{1}{m^{2}_{V_l}}I^{(b)}_{l\nu}, 
 \label{Ii1}
 \end{eqnarray}
with the integrals $I^{(a,b)}_{l\nu}$ being defined as 
\begin{eqnarray}
 I^{(a)}_{l\nu} & \equiv & \int\limits^{+\infty}_{-\infty}\frac{d^{4}q}{(2\pi)^{4}}\frac{p_{B\nu}}{q^{2} - m^{2}_{V_{l}} + i\epsilon}\frac{1}{(p_{B} - p_{E}-q)^{2} - m^{2}_{P_{l}} + i\epsilon},  \nonumber \\
I^{(b)}_{l\nu} & \equiv & \int\limits^{+\infty}_{-\infty}\frac{d^{4}q}{(2\pi)^{4}}\frac{(p_{B}\cdot{q})q_{\nu}}{q^{2} - m^{2}_{V_{l}} + i\epsilon}\frac{1}{(p_{B} - p_{E}-q)^{2} - m^{2}_{P_{l}} + i\epsilon} \label{Ii2}
\end{eqnarray}

The integrals in Eq.~(\ref{Ii1}) are calculated using the Passarino-Veltman reduction method for tensor integrals~\cite{Malabarba:2020grf,Passarino:1978jh}. Accordingly,  noticing that the integrals in Eq.~(\ref{Ii1}) are rank-$1$ tensors that depend on $p_{B\nu}$ and $p_{E\nu}$, the general Lorentz structure for $I^{(a,b)}_{l\nu}$ is given by 
\begin{eqnarray}
I^{(a,b)}_{l\nu} & = & a^{(a,b)}_{1l}p_{B\nu} + a^{(a,b)}_{2l}p_{E\nu}, 
\label{Ii3}
\end{eqnarray}
where $a^{(a,b)}_{1l}, a^{(a,b)}_{2l}$ are coefficients to be calculated. Using the preceding equations, we can write the amplitude in Eq.~(\ref{amplB1}) as follows,
\begin{eqnarray}
t^{(R)}& = &\left( A p^{\mu}_{B}\epsilon_{D^{*+\mu}} + B p^{\mu}_{E}\epsilon_{D^{*+\mu}}\right), 
    \label{ampSAB}
\end{eqnarray}
where 
\begin{eqnarray}
    A  &\equiv &  i\sum_{l=1}^8 (C^\text{ext}_0\omega^\text{ext}_l+C^\text{int}_{0}\omega^\text{int}_l) \bigg[ a^{(a)}_{1l} - \frac{a^{(b)}_{1l}}{m^{2}_{V_l}}\bigg]T_{l} \  , 
    \nonumber \\
    B  &\equiv&  i\sum_{l=1}^8 (C^\text{ext}_0\omega^\text{ext}_l+C^\text{int}_{0}\omega^\text{int}_l)\bigg[ a^{(a)}_{2l} - \frac{a^{(b)}_{2l}}{m^{2}_{V_l}}\bigg]T_{l} \ .
    \label{coeffAB}
\end{eqnarray}
Note that the above coefficients depend on the invariant mass of the interacting pair $D^{*+}\pi^-$ in the final state through, for example, the two-body $t$-matrix $T_l$ describing the transition $V_l P_l\to D^{*+}\pi^{-}$.

To determine the coefficients  $a^{(a,b)}_{1l}, a^{(a,b)}_{2l}$, we adopt the following strategy: the tensor integrals  $I^{(a,b)}_{l\nu}$ in Eq.~(\ref{Ii3}),  when contracted with the Lorentz structures $p_{B}^{\nu}$ and $p_{E}^{\nu}$, yields the following system of coupled equations:
\begin{equation}
   \left\{
\begin{matrix}
p^{\nu}_{B}\cdot I^{(r)}_{l\nu} & = & a^{(r)}_{1l}p^{2}_{B} & + & a^{(r)}_{2l}p_{B}\cdot p_{E}
\\
p^{\nu}_{E}\cdot I^{(r)}_{l\nu} & = & a^{(r)}_{1l}p_{B}\cdot p_{E} & + & a^{(r)}_{2l}p^{2}_{E}
\end{matrix}
\right.
\label{system1}
\end{equation}
where $r = a,b$. 
Therefore, the coefficients \textbf{$a^{(r)}_{1(2)l}$} can be written in terms of the scalar integrals $p^{\nu}_{B}\cdot I^{(r)}_{l\nu}$ and $p^{\nu}_{E}\cdot I^{(r)}_{l\nu}$ as
\begin{eqnarray}
    a^{(r)}_{1l} = \frac{p^{2}_{E}(p^{\nu}_{B}\cdot I^{(r)}_{l\nu}) - p_{E}\cdot p_{B}(p^{\nu}_{E}\cdot I^{(r)}_{l\nu})}{p^{2}_{E}p^{2}_{B} - (p_{E}\cdot p_{B})^2}, \nonumber \\
     a^{(r)}_{2l} = \frac{p^{2}_{B}(p^{\nu}_{E}\cdot I^{(r)}_{l\nu}) - p_{B}\cdot p_{E}(p^{\nu}_{B}\cdot I^{(r)}_{l\nu})}{p^{2}_{B}p^{2}_{E} - (p_{B}\cdot p_{E})^2}.
     \label{coeff1}
\end{eqnarray}

To determine the quantities $p^{\nu}_{B}\cdot I^{(r)}_{l\nu}$ and $p^{\nu}_{E}\cdot I^{(r)}_{l\nu}$ appearing in Eq.~(\ref{coeff1}), we use Eq.~(\ref{Ii2}), contract them with either $p^\nu_B$ or $p^\nu_E$, and write explicitly the dependence on the $q^0$ variable of the integrand. In this way,  the integration over $q^0$ can be done analytically by using Cauchy’s theorem. To do this, we define   
\begin{eqnarray}
   \Pi^{(n)}_{l}&\equiv& \int\limits^{+\infty}_{-\infty}\frac{dq^0}{(2\pi)}  (q^{0})^n \, D_{l},
\label{Pin2}
\end{eqnarray}
where
\begin{align}
   D_{l}&\equiv\frac{1}{[q^{0} - \omega_{V_l} + i\epsilon_1][q^{0} + \omega_{V_l}- i\epsilon_1]}\nonumber\\
   &\quad\times \frac{1}{[q^{0} +p_E^0 - m_B  -\omega_{P_{l}} + i\epsilon_2][q^{0} + p_E^0 - m_B + \omega_{P_{l}} - i\epsilon_2]}, 
\label{Pin3}
\end{align}
with 
\begin{eqnarray}
  \omega_{V_l} & = & \sqrt{\vec{q}^2 + m_{V_l}^2}, 
   \\
    \omega_{P_l} & = &  \sqrt{(\vec{q} + \vec{p_E})^2 + m_{P_l}^2}.
\label{Pin3b}
\end{eqnarray}
We should mention here that Eq.~(\ref{Pin3}) has been written in the rest frame of the decaying particle, which we consider from now onward.

By calculating the residues related to the poles at $\omega_{V_l}-i\epsilon_1$ and $m_B - p_E^0 +\omega_{P_l}-i\epsilon_2$, we can determine an analytical expression for Eq.~(\ref{Pin2}) and the structures $p^{\nu}_{B}\cdot I^{(r)}_{l\nu}$ and $p^{\nu}_{E}\cdot I^{(r)}_{l\nu}$ appearing in Eq.~(\ref{coeff1}) can be written in terms of $\Pi^{(n)}_l$ as:
\begin{eqnarray}
p^{\nu}_{B}\cdot I^{(a)}_{l\nu} & = & \int\limits^{+\infty}_{-\infty}\frac{d^{3}q}{(2\pi)^{3}}m_B^2\Pi^{(0)}_{l}, 
\nonumber\\
p^{\nu}_{B}\cdot I^{(b)}_{l\nu} & = & \int\limits^{+\infty}_{-\infty}\frac{d^{3}q}{(2\pi)^{3}}m_B^2\Pi^{(2)}_{l},
\nonumber \\
p^{\nu}_{E}\cdot I^{(a)}_{l\nu} & = & \int\limits^{+\infty}_{-\infty} \frac{d^{3}q}{(2\pi)^{3}}p^{0}_{E}m_{B}\Pi^{(0)}_{l},
\nonumber\\
p^{\nu}_{E}\cdot I^{(b)}_{l\nu} & = & \int\limits^{+\infty}_{-\infty} \frac{d^{3}q}{(2\pi)^{3}}p^{0}_{E}m_B(\,\Pi^{(2)}_{l} - \vec{q}^{\ 2}\,\Pi^{(0)}_{l}).
\label{Pin1}
\end{eqnarray}
The integration on $d^3q=d\text{cos}\theta d\phi d|\vec{q}||\vec{q}|^2$ in the preceding equations is done numerically, with the integral in $d|\vec{q}|$ being regularized with a form factor. Here, we consider two different types of form factors to estimate uncertainties: a monopole, $F_M(|\vec{q}|),$ and a Gaussian, $F_G(|\vec{q}|)$, form, with
\begin{align}
    F_M(|\vec{q}|)=\frac{\Lambda^2_M}{\Lambda^2_M+|\vec{q}|^2},\nonumber\\
    F_G(|\vec{q}|)=e^{-|\vec{q}|^2/\Lambda_G^2}.\label{FF}
\end{align}
The $\Lambda_M$, $\Lambda_G$ in Eq.~(\ref{FF}) are of the order of $\sim 1000$ MeV, and we vary them in the range $600-1100$ MeV to estimate uncertainties in the model.

\bibliographystyle{apsrev4-1}
\bibliography{D1refs}

\end{document}